\documentclass[twocolumn,showpacs,floatfix,amsfonts]{revtex4}
\usepackage{graphicx,graphics,color,epsfig}
\usepackage{bm}
\usepackage{amsmath}
\usepackage{amssymb}

\begin{document}

\title{Reentrant stability of BEC standing wave patterns}
\author{Ryan M. Kalas$^{1,2}$, Dmitry Solenov$^{1,2}$, and Eddy Timmermans$^{1}$}
\affiliation{$^{1}$ Center for Nonlinear Studies (CNLS),
Los Alamos National Laboratory, Los Alamos, NM 87545, USA \\
$^{2}$T-4, Theoretical Division, Los Alamos National
Laboratory, Los Alamos, NM 87545}

\date{October 30, 2009}

\begin{abstract}

We describe standing wave patterns induced by an attractive finite-ranged external 
potential inside a large Bose-Einstein Condensate (BEC).  As the potential depth increases, 
the time independent Gross-Pitaevskii equation develops pairs of solutions that have 
nodes in their wavefunction.  We elucidate the nature of these states and study their
dynamical stability.  Although we study the problem in a two-dimensional BEC subject
to a cylindrically symmetric square well potential of a radius that is comparable to the 
coherence length of the BEC, our analysis reveals general trends, valid in two and
three dimensions, independent of the symmetry of the localized potential well, and
suggestive of the behavior in general, short- and large-range potentials.
One set of nodal BEC wavefunctions resembles the single particle $n$ node bound state 
wavefunction of the potential well, the other wavefunctions resemble the $n-1$ node bound-state
wavefunction with a kink state pinned by the potential.  The second state, though 
corresponding to the lower free energy value of the pair of $n$ node BEC states,  is always
unstable, whereas the first can be dynamically stable in intervals of the potential
well depth, implying that the standing wave BEC can evolve from a dynamically unstable
to stable, and back to unstable status as the potential well is adiabatically deepened,
a phenomenon that we refer to as ``reentrant dynamical stability''.

\end{abstract}
\maketitle

\section{Introduction}

\begin{figure}[htb]
\center{\includegraphics[width=3.0in]{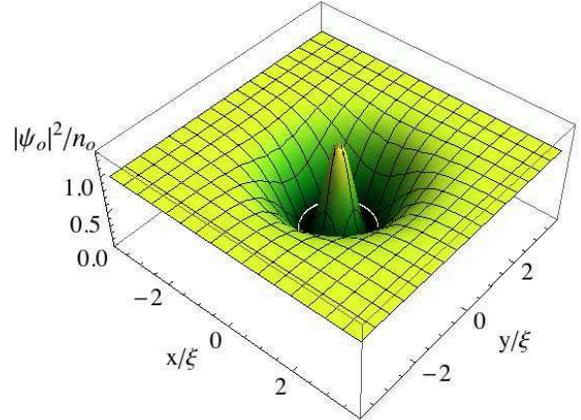}}  
\caption{ \label{figintro} Example of an atomic nodal BEC density (well radius $b=1.0\,\xi$, well depth $\eta=1.8$).
This case happens to be in a reentrant stable regime.  }
\end{figure}

As a superfluid, the dilute gas Bose-Einstein condensate (BEC) is a coherent 
quantum system and behaves in many respects like a single-particle quantum wave.  
Stationary single particle bound states have wave functions with nodes, 
giving standing wave patterns. In a BEC, nodal structures, such as the one pictured
in Fig.~\ref{figintro}, could be imaged and manipulated directly since the experimental 
BEC wave functions extend over tens of microns.  Can BECs support such 
patterns as long-lived structures?  A fundamental difference between single 
particle and BEC wave dynamics is the nonlinearity of the BEC evolution that 
stems from the inter-particle interactions.  This nonlinearity leads to 
stationary standing wave solutions such as kink states or dark solitons that have
been created in BEC experiments \cite{Denschlag,Hau}.  However, the same 
nonlinearity also makes the kink states unstable in two and three dimensions
 \cite{Josserand,muryshev,muryshev2}, as has been observed  in cold atom 
experiments \cite{Anderson}.  The work of this paper was motivated by the 
basic question: when and how can a BEC support a long-lived standing wave
induced by a local potential well in dimensions higher than one?  The standing
wave gives a stationary interference pattern in the full single-particle density 
that vanishes \cite{remark1} on the surfaces at which the stationary BEC wavefunction 
changes sign.  The existence of such structures may, as we discuss below, 
impact fundamental science phenomena and cold atom applications.  

We analyze the nodal BEC wavefunctions in a two-dimensional (pancake shaped) 
BEC; although the trends uncovered by our analysis, suitably generalized, are valid in higher 
dimensions. The 2D trap geometry, realized in cold atom experiments \cite{Dalibard,Malcolm}, 
offers unprecedented prospects for ``wavefunction 
engineering'': focused laser-beams can, in principle, access and image each point 
of the BEC wavefunction.  We consider a large, dilute 2D BEC in a homogeneous
potential that is subjected to a superimposed attractive potential well of cylindrical 
symmetry with a radius comparable to the coherence length of the surrounding BEC.  
We study the existence of cylindrically symmetric standing wave solutions of the 
time-independent Gross-Pitaevskii equation, elucidating their nature and investigating 
their dynamical stability.

Our analysis reveals general trends: as the depth of the finite-ranged potential 
well increases near the resonance value (though still above) at which a 
single-particle bound state of $n$ nodes forms,  a pair of $n$ node standing wave solutions 
to the time independent Gross-Pitaevskii equation appears,
as found in Ref.~\cite{massignan}.  Upon further deepening of the well, the last 
node of one of the two new BEC wavefunctions moves inside the well 
and the corresponding wavefunction resembles that of the single 
particle bound state.  We refer to this solution as the ``atomic nodal BEC'' state.  
At the same well depth, the other $n$ node BEC-wavefunction resembles the 
wavefunction of the $n-1$ node atomic bound state with an extra soliton pinned 
near the potential edge and we refer to it as the ``nonlinear nodal BEC'' solution.  Of the two 
$n$ node BEC wavefunctions, the nonlinear nodal BEC state has the lower free energy, 
but we find that it is always dynamically unstable.  If the atomic nodal BEC wavefunction 
varies on a length scale comparable to or greater than the healing length of the surrounding BEC, 
then the atomic BEC solution is also unstable when it first forms.  Upon further deepening 
of the well, the stationary atomic nodal BEC wavefunction can become stable, then go 
unstable again as the well depth is further increased---a phenomenon that we refer
to as ``reentrant dynamical stability''.  We show that the stability analysis
can be reduced to the simpler study of the spectrum of two distinct 
single-particle-Hamiltonian-like operators.  

The size of the stability islands, in terms of the range of well depths, 
for the atomic nodal BEC standing wave depends
sensitively on the ratio of the potential range to the BEC coherence length:
the more the potential range exceeds the coherence length, the broader the
regions of instability become.  If the potential well is confined to a spatial
region much smaller than the BEC coherence length, the regions of instability 
are confined to narrow intervals of the well depth and the time scale at which 
the instability sets in increases markedly. 

The existence of stable, localized BEC standing waves 
could impact many areas of cold atom physics.  In cold atom 
traps standing waves could play a role in the physics of localized objects moving through 
a BEC \cite{pitaevskii}.  If the object interacts strongly and attractively with BEC bosons, the 
dynamics of object acceleration may bring the BEC into an unstable nodal 
state, disturbing the system and possibly providing another mechanism of invalidating
the picture of superfluid, dissipationless motion \cite{david}.  A class of BEC objects of 
particular, fundamental interest consists of a BEC with self-localized polaron-like impurity atoms 
\cite{eddy,ryan}.  For sufficiently strong impurity-boson attractions, a standing 
wave BEC configuration combined with a localized, nodal impurity wavefunction
may give more complex, localized excited state polaron objects.  Another
class of BEC objects that was proposed \cite{cote} consists of mesoscopic
ultra-cold molecules.  Localized potentials of ions embedded in the BEC 
may capture cold atoms in micron-sized orbits.  If the BEC can be engineered 
to be in an atomic nodal BEC state,
a gentle lowering of the boson-boson interactions can bring the system
into a true molecular state, assuming that no instabilities are encountered in
the adiabatic dynamics.  Finally, we remark that long-lived, localized standing wave BEC patterns 
may resolve the challenge of observing the BEC response to external 
perturbations.  For instance, a ring BEC can sense rotation: as the trap
rotates in the plane of the ring, the BEC cannot follow as such motion
would imply vorticity.  Hence, in the rotating trap frame, the BEC flows.
How can we observe that flow?  If a local dip in the ring potential
supports a standing wave BEC with fringes (across which the BEC cannot flow),
the rotation would shift the position of the fringes, perhaps destroying the standing wave.

For a more whimsical application of standing wave nodal BEC states, one could look to 
the hypothesized dark matter BECs.
If, as has been speculated,
the dark matter haloes that surround most galaxies are scalar BECs of
ultra-light boson particles \cite{sin,bohmer,sikivie}, steep gravitational potentials 
may support standing wave dark matter BEC density patterns. 

The relevance of any of the prospects for standing wave BECs hinge
on their stability.  In this manuscript we study the standing
wave patterns and their stability in a linear stability analysis (dynamical stability).
The paper is organized as follows:  Section II describes the multiple, stationary 
solutions to the Gross-Piteavskii equation in the presence of
an attractive potential well of finite radius.  We describe the linear stability analysis
in Section III and we analyze the dynamical stability of the standing wave solutions 
in Section IV.  In Section V, we describe a general analysis based on
the study of the spectrum of effective single-particle Hamiltonians.  Section VI concludes.

\section{Standing Wave BEC Wavefunctions}

We consider a stationary, dilute two-dimensional BEC of $N$ bosons, each of mass $m_{b}$.  
Such systems can be realized with cold atom harmonic oscillator traps by increasing 
one of the the trap frequencies $\omega_{z}$ to ensure that the corresponding energy 
significantly exceeds the chemical potential $\mu$ of the BEC system, 
$\hbar \omega_{z} \gg \mu$.  The bosons interact mutually via a short-range interaction 
described by a contact interaction potential, $v({\bf r}-{\bf r}')=g \delta({\bf r}-{\bf r}')$,
where ${\bf r}$ is a 2D vector and $\delta({\bf r})$ the 2D delta function.
In the dilute regime, the interaction strength $g$ relates to the scattering length
$a$ that characterizes low energy scattering in three dimensions and to the
ground state extent $l_{z}$ of the $\omega_{z}$ frequency, $l_{z}
=\sqrt{\hbar/m_{b} \omega_{z}}$, by the relation $g=(\sqrt{8\pi}\hbar^{2}/m_{b}) (a/l_{z})$ 
\cite{petrov,blochRMP}.  
Here we assume that the $N$ bosons are contained by a 2D-box potential of
macroscopic area $\Omega$, corresponding to an average density $n_{0}=N/\Omega$.
The BEC is described by a wavefunction (or order parameter) $\psi_{0}({\bf r})$ that
is normalized by requiring $\int_{\Omega} d^{2} r |\psi_{0}({\bf r})|^{2} = N$.  In addition,
the BEC experiences an attractive external potential $V_{\rm ext}({\bf r})$ of finite
range.  The wavefunction satisfies the time-independent Gross-Pitaevskii equation
\begin{equation}
\label{gp}
\Big{(} -\frac{\hbar^2}{2m_{b}}\nabla^2 + V_{\rm ext}(r) + g|\psi_{\rm o}|^2 \Big{)}\psi_{\rm o} =\mu \psi_{\rm o} {\rm.}
\end{equation}
Far from the center of the external potential, the wavefunction tends to a constant value,
$\psi_{o} \rightarrow \sqrt{n_{o}}$, so that the chemical potential is related to the density
by $\mu=gn_{o}$.  The natural length scale of the problem is the coherence length of the BEC,
$\xi$, defined by $\hbar^{2}/m_{b} \xi^{2}=\mu$.  Scaling the energy by $\mu$, the length by
$\xi$, ${\bf x}={\bf r}/\xi$, and the density by $n_{o}$, so that $\phi_o=\psi_o/\sqrt{n_{o}}$,
we obtain the dimensionless form of the Gross-Pitaevskii equation,
\begin{equation}
\label{gp2}
\Big{(} -\frac{\nabla_{{\bf x}}^2}{2} + U_{\rm o}({\bf x})+|\phi_{\rm o}|^2 \Big{)}\phi_{\rm o} 
   = \phi_{\rm o}   {\rm ,}
\end{equation}
where $U_o({\bf x})$ denotes the scaled external potential $U_{\rm o}({\bf x})
= V_{\rm ext}({\bf r}={\bf x}\,\xi)/\mu$, and the large distance boundary condition
becomes $\lim_{x\rightarrow \infty} \phi_{\rm o} ({\bf x}) = 1$.

We perform the calculations for an attractive external potential that has a square well 
shape of radius $b$
and depth  $\pi^2 \hbar^2 \eta^2 / 8 m_{b} b^2$.  In three dimensions the resonances 
for noninteracting particles in this potential would occur when $\eta$ is an odd integer;
while in 2D the noninteracting resonances occur at $\eta\approx2.45$, $4.48$, $\ldots$,
relating to solutions of a transcendental equation involving Bessel functions. 
Eq.~(\ref{gp2}) then reads
\begin{equation}
\Big{(} -\frac{\nabla^2}{2} -\frac{\pi^2\eta^2}{ 8 (b/\xi)^2}\Theta(b/\xi -x)+
|\phi_{\rm o}|^2 \Big{)}\phi_{\rm o} 
   = \phi_{\rm o}   {\rm ,}
\label{gp3}
\end{equation}
where $\Theta$ is the unit step function.
For a given well radius $b/\xi$, we vary the dimensionless well depth $\eta$ and solve Eq.~(\ref{gp3}).

\begin{figure}[htb]
\center{\includegraphics[width=3.3in]{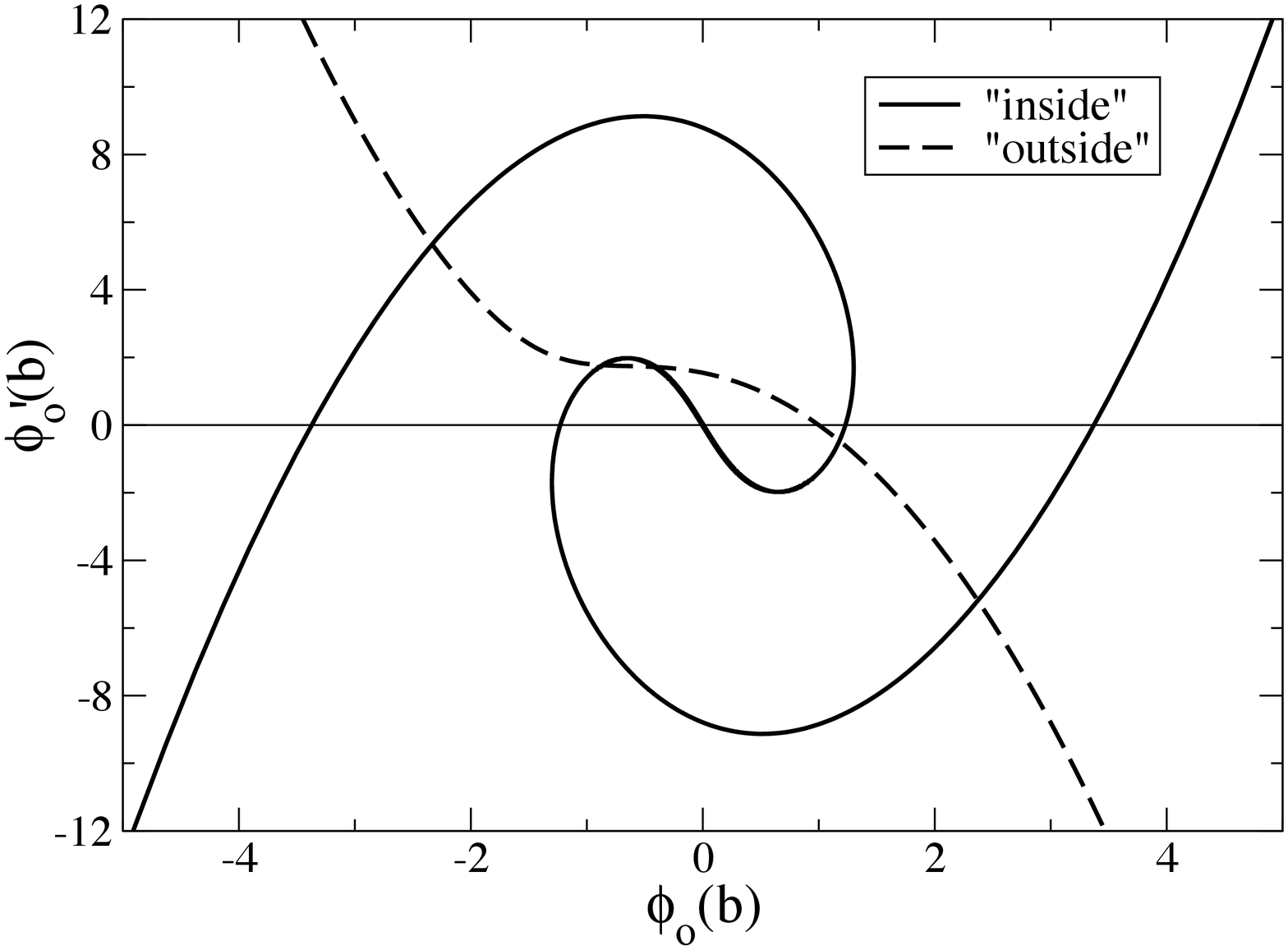}}  
\caption{ \label{figmatch} Solving Eq.~(\ref{gp3}) graphically, matching inner (solid) and 
outer (dashed) integrations at the potential edge, $x_m=b/\xi$, by locating the crossing points 
in the ($\phi_{\rm o}'(b/\xi),\phi_{\rm o}(b/\xi)$) plane.
The example shown, for a well radius of $b=1.0\,\xi$ and well depth $\eta=2.9$, gives five 
crossing points and thus five solutions to the GP equation, Eq.~(\ref{gp3}).}
\end{figure}

To integrate Eq.~(\ref{gp3}) subject to the proper boundry conditions, we use a scheme 
introduced in Ref.~\cite{massignan} to construct the time-independent BEC wavefunctions in the 
ion-neutral atom polarization potential.  
The boundary conditions introduce parameters $A_{1}$ and $A_{2}$:  at the origin, $\phi_{\rm o}'(x=0)=0$ and 
$\phi_{\rm o}(x=0)=A_{1}$, whereas far from the localized potential, 
$\phi_{\rm o}(x) \approx 1 + A_2 \exp(-2x)$.
We pick the radius of the potential well, $x_{m}=b/\xi$, as the matching 
point for the inward and outward integration procedure.  
Integrating inward from large $x$-values for different choices of $A_{2}$, we find 
the wavefunction and its derivative at the matching point,  $\left(\phi_{\rm o}(x_{m}), 
\phi_{\rm o}'(x_{m})\right)$, tracing out a parametric curve that is independent of the 
potential (since the equation is integrated in the region where $U_{\rm o}({\bf x})=0$ for  $x>x_{m}$).  
Similarly, the outward integration from $x=0$ for different $A_{1}$ values yields a 
second $\left(\phi_{\rm o}(x_{m}), \phi_{\rm o}'(x_{m})\right)$ curve, this one dependent 
on the potential.  As the values of the wavefunction and its derivative must match at $x_{m}$, 
the intersections of both curves (and the corresponding $A_{1}$ and $A_{2}$ parameter values) 
determine the stationary solutions.  Fig.~\ref{figmatch} shows the curves for well radius $b=1.0\,\xi$ 
and well depth $\eta = 2.9$.  These parameters result in five crossings and thus five solutions.  
As mentioned, the ``outside'' curve doesn't depend on well depth $\eta$.  When $\eta$ 
is small, there is only a single crossing of the outside curve with the ``inside'' curve, corresponding to a
single solution---the familiar nodeless BEC. As the well depth $\eta$ increases, the 
inside curve starts to swirl and pinches through the outside curve, generating
a pair of new solutions.   When the well depth has reached the $\eta=2.9$ value shown in 
Fig.~\ref{figmatch}, the inside curve pinches once more through the outside curve
near $\phi_{\rm o}(b)\approx -0.8$, generating two additional time-independent
Gross-Pitaveskii solutions.

The new solutions differ qualitatively from the nodeless BEC solution.  As $\eta$
increases, each one of the new pair of BEC wavefunctions has one more radial node than
the number of nodes of the previous pair.  At the well depth where the new pair of solutions first exists, 
the two new solutions are degenerate and they have the last node located slightly beyond the well radius.  
For a well radius of $b=1.0\,\xi$, we find the critical well depth for the one node solutions 
\begin{figure}[htb]
\center{\includegraphics[width=3.1in]{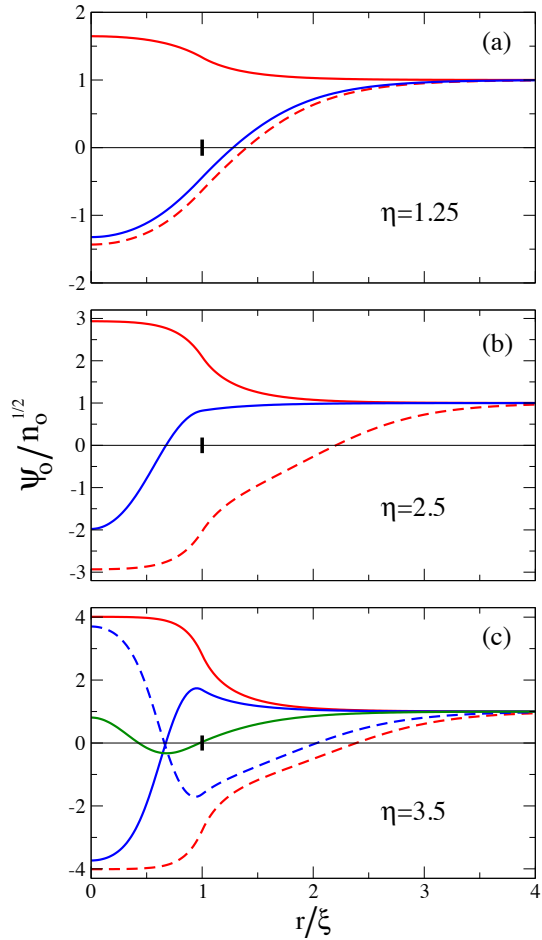}}  
\caption{ \label{figwf} Wavefunctions $\psi_{\rm o}/\sqrt{n_{\rm o}} = \phi_{\rm o}$ 
solving Eq.~(\ref{gp3}) for a well radius $b=1.0\,\xi$ (marked as a hash on axis) 
and well depths $\eta$ of (a) $1.25$, (b) $2.5$, and (c) $3.5$. As the well is made
deeper, multiple solutions are found.}
\end{figure} 
is $\eta\approx1.2444$.  In Fig.~\ref{figwf}(a) for $\eta=1.25$, the
two new solutions have visibly split.  As the well depth increases, the two nodal
solutions continue to differentiate: the node of one moves inside the potential
well as the well depth increases, while the the node of the other solution
moves outside the potential well as depicted in Fig.~\ref{figwf}(b) for
well depth $\eta=2.5$.  As the well depth increases further, 
two new solutions, each with two nodes, appear near $\eta\approx2.86$;
and we show the five solutions for $\eta=3.5$ in Fig.~\ref{figwf}(c).

\begin{figure}[htb]
\center{\includegraphics[width=3.3in]{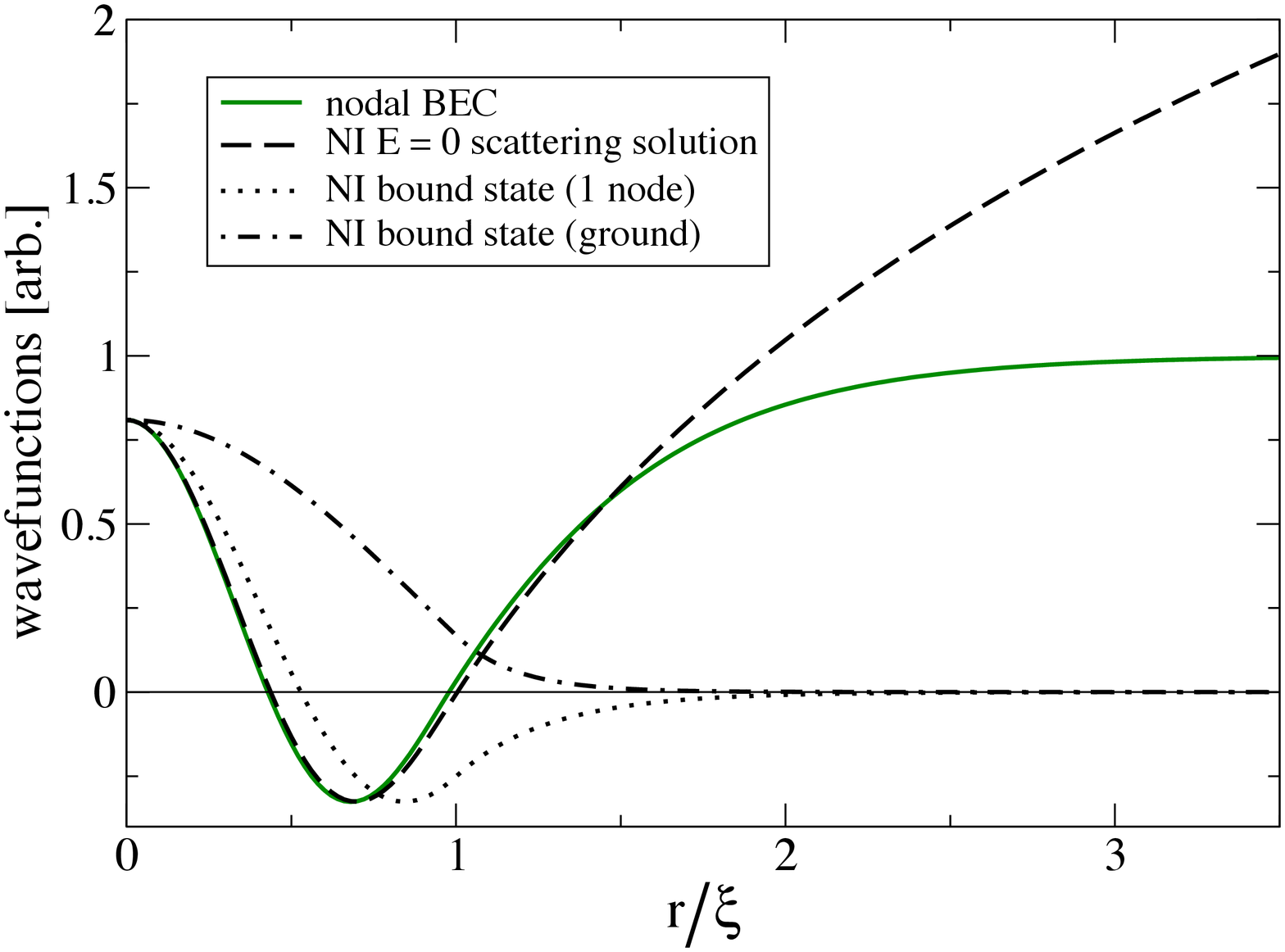}} 
\caption{ \label{fig_cp_phi}  For a well depth of $\eta=3.5$, the noninteracting (NI)
bound states with zero and one node and the NI zero energy scattering solution
with two nodes are plotted to compare to the interacting solutions 
with the same well depth in Fig.~\ref{figwf}(c). In particular, the (green) nodal
BEC solution with two nodes is similar to the zero energy scattering wavefunction.}
\end{figure}

We clarify the nature of the different nodal solutions. 
In accordance with Levinson's theorem, the deepening of the
potential well produces bound states in the single particle
wavefunction description.  Each time the zero energy scattering 
phase goes through a resonance (increasing by $\pi$), a new
bound state forms with one more node than the previously
formed bound state.  We note that one class of the time-independent 
Gross-Pitaevskii solutions---the solution in which the last node moves
inward as the potential depth increases---starts to resemble the noninteracting
(NI) zero energy wavefunction.
In Fig.~\ref{figwf}(c), the solid line curves show three BEC wavefunctions with zero,
one, and two nodes (the dashed curve wavefunctions will be discussed below).
Compare these wavefunctions to the NI wavefunction plotted in Fig.~\ref{fig_cp_phi}
for the same well depth.  The NI bound states with zero and one node
exhibit a qualitative similarity to the BEC wavefunctions with zero and one nodes, except that the
former vanish at large distance whereas the latter tend to a constant 
value (Figs.~\ref{fig_cp_phi} and \ref{figwf}(c)).
Moreover, the NI zero-energy scattering solution is practically
identical in form to the nodal BEC wavefunction with two nodes inside the well,
except for the asymptotic behavior outside the well---see the solid
and dashed lines in Fig.~\ref{fig_cp_phi} \cite{notenodej}.  Inside the 
well the nonlinearity in Eq.~(\ref{gp2}) has little effect on these nodal solutions 
and the nodal BEC wavefunction strongly resembles single-particle orbitals.
To stress the similarity, we will refer to the BEC solutions discussed above 
(the solid lines in Fig.~\ref{figwf}) as the ``atomic nodal BEC'' solutions.
 
\begin{figure}[htb]
\center{\includegraphics[width=3.0in]{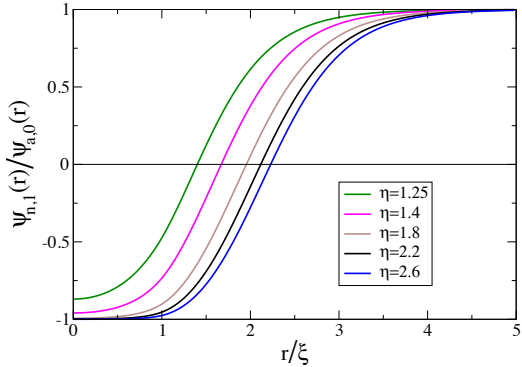}} 
\caption{ \label{fig_psi1onpsi0} Plots of $\psi_{{\rm n},1}(r)/\psi_{{\rm a},0}(r)$ 
for various well depths; the quotient takes on the hyperbolic tangent form discussed
below Eq.~(\ref{nonlinearwf}), indicating that the solution $\psi_{{\rm n},1}$
can be interpreted as a kink added to the solution $\psi_{{\rm a},0}$. }
\end{figure}

We show that the wavefunction of the other solutions (with a node that moves further 
away from the well edge as the well-depth increases), plotted by the red and blue 
dashed lines in Fig.~\ref{figwf}(b-c), 
are related to the nonlinear physics of solitons.   Note that the curves of the same color 
in Fig.~\ref{figwf}(b-c) have the same peak magnitude and would have similar shapes 
(except for the outermost node). 
We label the solid line solutions as $\psi_{{\rm a},i}(r)$ 
and the dashed solutions as $\psi_{{\rm n},i}(r)$, representing the type of solution
by the first (``atomic nodal BEC'' and ``nonlinear nodal BEC'') and the number of nodes by the second
subscript.   The ``nonlinear nodal BEC'' solutions, when they exist, are well approximated as 
\begin{equation}
\label{nonlinearwf}
\psi_{{\rm n},i}(r) \approx \psi_{{\rm a},i-1}(r)\chi(r-R),
\end{equation}
where $\chi(r)=\tanh(r/\xi)$ and $R$ is the position of the outermost node. 
This hyperbolic tangent is the familiar nonlinear kink or soliton solution:
the $\psi_{n,i}$ solution has a soliton that is trapped or pinned 
outside the potential well.  We illustrate this point graphically in 
Fig.~\ref{fig_psi1onpsi0} by plotting $\psi_{{\rm n},1}(r)/\psi_{{\rm a},0}(r)$ for 
different well depths $\eta$.  Note that the quotient follows the hyperbolic 
tangent shape.
The red dashed (1 node) wavefunction in Fig.~\ref{figwf} can be interpreted as 
the red solid (nodeless) wavefunction with a kink superimposed on it.  Similarly, the blue dashed
(2 node) wavefunction can be viewed as the blue solid wavefunction with a kink.  For deeper wells,
this pattern repeats, e.g., for deeper wells the green (2 node) wavefunction develops
a kinked partner with 3 nodes.

\begin{figure}[htb]
\center{\includegraphics[width=3.3in]{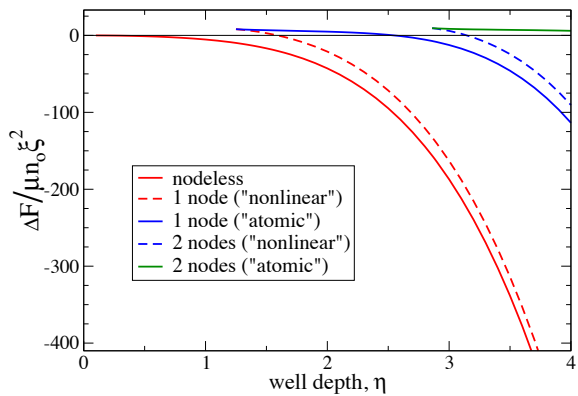}}   
\caption{ \label{figdF} Free energy difference between the system with the potential well present
and the homogeneous system without well, as given by Eq.~\ref{dF}, for a well radius of $b=1.0\,\xi$.
The colors/dashings correspond to those in Fig.~\ref{figwf}. }
\end{figure}

Which solution has the highest energy?  By fixing the density far away 
from the localized potential we fix the chemical potential so that we have to determine
the free energy, $F = H - \mu N$, where $H$ is the Hamiltonian and $N=\int {\rm d}^2r \, |\psi_o(r)|^2$.  
In fact, we calculate the difference between the free-energy of the $\phi_{\rm o}$ solution and
that of the homogeneous system of the same density (without external potential), 
\begin{equation}
\label{dF}
\frac{\Delta F}{\mu n_o \xi^2} = \int {\rm d}^2x \, \Big{[} \frac{|\nabla \phi_{\rm o}|^2}{2} 
+ U_{\rm o}|\phi_{\rm o}|^2 + \frac{1}{2}( |\phi_{\rm o}|^4 +1 ) -\phi_{\rm o}^2 \Big{]}.
\end{equation}
In Fig.~\ref{figdF}, we plot $\Delta F$ vs. the well depth $\eta$ for the nodeless 
and the two nodal solutions for  a well radius of $b=1.0 \,\xi$.  
For ease of comparison, the colors/dashings of the free energies in Fig.~\ref{figdF} are
the same as those of the wavefunctions in Fig.~\ref{figwf}.

As with noninteracting systems, the nodeless solution (red line) has the lowest free energy.
The introduction of an attractive well smoothly lowers $\Delta F$ from $0$
to negative values as the well is turned on.
At the well depths where the nodal solutions first appear,
their $\Delta F$ is positive for a range of well depths
before becoming negative \cite{notedfpos}.
The nonlinear pinned soliton solutions of n nodes (depicted by the red and blue dashed 
lines in Fig.~\ref{figdF}) have a $\Delta F$ value that differs from that of the atomic $n-1$ node solutions 
by a nearly constant value---the energy cost of creating the kink. 

The solutions with nodes are obviously not the ground state of the system, but 
are they metastable?  In the next sections, we undertake a
stability analysis to investigate whether they might be long-lived.

\section{Linear Stability Analysis: Formalism}

In this section and the following we test whether or not the
$\phi_{\rm o}$ function is stable with respect to small perturbations.  We use
linear stability analysis, which investigates the response of the system 
to a weak perturbation caused, for instance, by a modulation of the potential 
around the time-independent external potential $U_{\rm o}$,
$U({\bf x},t)=U_{\rm o}({\bf x})+\delta U({\bf x},t)$.  The response of the system
is described by the time-dependent Gross-Pitaevskii equation,
\begin{equation}
i \frac{\partial \phi({\bf x},t)}{\partial t} 
= \left[ - \frac{\nabla^{2}}{2} + |\phi({\bf x},t)|^{2} + U({\bf x},t) \right] \phi({\bf x},t).
\label{tgp}
\end{equation}
The time-independent solution to the equation with $U=U_{\rm{o}}$ evolves as
$\phi({\bf x},t)=\exp(-it) \phi_{\rm o}({\bf x})$.  The  system's small amplitude response 
\begin{equation}
\phi({\bf x},t) \approx \exp(-it) \left[ \phi_{\rm o} ({\bf x}) 
+ \delta \phi ({\bf x},t) \right] ,
\label{fluct}
\end{equation}
to the potential perturbation $U({\bf x},t) = U_{\rm o} ({\bf x})
+ \delta U({\bf x},t)$ evolves according to the linearized Gross-Pitaevskii equation
for the $\delta \phi$ evolution,
\begin{equation}
i \frac{\partial \delta \phi}{\partial t}
= \hat{h}^{-} \delta \phi
+ \phi_{\rm o} \left( \phi_{\rm o}^{\ast} \delta \phi
+ \phi_{\rm o} \delta \phi^{\ast} \right) + \phi_{\rm o} \delta U ,
\label{lgp}
\end{equation}
where the $\hat{h}^{-}$ operator,
\begin{equation}
\hat{h}^{-} = - \frac{\nabla^{2}}{2} + U_{\rm o}
+ \left( |\phi_{\rm o}|^{2} -1 \right),
\label{h-}
\end{equation}
represents the Hartree-Hamiltonian of the $\phi_{\rm o}$ system.

Note that the linearized Gross-Pitaevskii equation, Eq.~(\ref{lgp}),
couples $\delta \phi$ and $\delta \phi^{\ast}$, mixing wavefunction fluctuations
and their complex conjugates.   To describe the coupling, we separate out 
the positive from the negative frequency components in the Fourier transform,
\begin{eqnarray}
\delta \phi ({\bf x},\omega) &=& \int_{-\infty}^{\infty} dt e^{i \omega t}
\delta \phi({\bf x},t) ;
\nonumber \\
\delta \phi({\bf x},t) &=& 
\frac{1}{2\pi} \int_{-\infty}^{\infty} d \omega e^{- i \omega t} \delta
\phi({\bf x},\omega) \; .
\label{ft}
\end{eqnarray}
We define the positive, $\delta \phi_{p}$ and negative, $\delta \phi_{n}$, frequency
amplitude functions to have non-zero-values for $\omega >0$,
\begin{eqnarray}
\delta \phi_{p}({\bf x},\omega) &=& \delta \phi({\bf x},\omega) 
\; \; \; \; {\rm if} \; \; \omega > 0 ;
\nonumber \\
&=& 0  \; \; \; \;  {\rm if} \; \; \omega < 0 ;
\nonumber \\
\delta \phi_{n}({\bf x},\omega) &=& \delta \phi^{\ast}({\bf x}, - \omega)
 \; \; \; \; {\rm if} \; \; \omega > 0 ;
 \nonumber \\
 &=& 0 \; \; \; \;  {\rm if}\; \; \omega < 0 ;
\label{defpn}
\end{eqnarray}
so that
\begin{equation}
\delta \phi({\bf x},\omega) = \delta \phi_{p}({\bf x},\omega)
+ \delta \phi_n^{\ast} ({\bf x},-\omega)  .
\label{deltaphi}
\end{equation}
If the time-dependent function $\delta U({\bf x},t)$ is real-valued, 
then the negative and positive frequency amplitudes are identical,
$\delta U_{n}({\bf x},\omega) = \delta U_{p}({\bf x},\omega)$,
where $\delta U_{n}$ and $\delta U_{p}$ are defined as in
Eq.~(\ref{defpn}).  
In the time-domain, 
\begin{eqnarray}
\hspace{-0.2in}\delta \phi({\bf x},t) =& \int_{0}^{\infty} d \omega \; e^{-i \omega t} \;
\delta \phi_{p} ({\bf x},\omega)
\nonumber \\
&+ \int_{0}^{\infty} d \omega \; e^{i \omega t} \; 
\delta \phi^{\ast}_{n} ({\bf x},\omega) 
\nonumber \\
=& \delta \phi_{p}({\bf x},t) + \delta \phi^{\ast}_{n}({\bf x},t) \; \; ,
\label{phipn}
\end{eqnarray}
where, in accordance with the Fourier transformation,
\begin{equation}
\delta \phi_{p(n)} ({\bf x},t) = \frac{1}{2\pi} \int_{- \infty}^{+\infty} d \omega \; 
e^{-i \omega t} \; \delta \phi_{p(n)} ({\bf x},\omega) ,
\end{equation}
and we remember that $\delta \phi_{p(n)} ({\bf x},\omega) = 0 $ if $\omega < 0$.

Inserting Eq.~(\ref{phipn}) into the linearized Gross-Pitaevskii equation of
Eq.~(\ref{lgp}), we identify the $e^{-i\omega t}$ and the $e^{i\omega t}$ components.
Taking the negative of the complex conjugate of the $e^{i\omega t}$ components,
we obtain
\begin{eqnarray}
\omega \delta \phi_{p} &=& \left[ \hat{h}^{-} + |\phi_{\rm o}|^{2} \right]
\delta \phi_{p} + \phi_{\rm o}^{2} \delta \phi_{n} + \phi_{\rm o} \delta U_{p};
\nonumber \\
\omega \delta \phi_{n} &=& - \phi^{2 \ast}_{\rm o} \delta \phi_{p}
- \left[ \hat{h}^{-} + |\phi_{\rm o}|^{2} \right] \delta \phi_{n} 
- \phi^{\ast}_{\rm o} \delta U_{n}.
\label{lgpo}
\end{eqnarray}
Multiplying by $e^{-i \omega t}$ and integrating over $(2 \pi)^{-1} \int d \omega$, we
obtain the time-domain version,
\begin{equation}
i \frac{\partial}{\partial t}
\left(
\begin{array}
[c]{l}
\delta \phi_{p} \\
\delta \phi_{n}
\end{array}
\right)
=
\hat{\cal L}
\left(
\begin{array}
[c]{l}
\delta \phi_{p} \\
\delta \phi_{n}
\end{array} 
\right)
+ 
\left(
\begin{array}
[c]{l}
\; \; \; \phi_{\rm o}  \delta U_{p} \\
-\phi^{\ast}_{\rm o} \delta U_{n}
\end{array}
\right),
\label{lgp2}
\end{equation}
where we have introduced the $\hat{\cal L}$ operator used in Ref.~\cite{castin1,castin2},
\begin{align}
\hat{\cal L} =
\left( 
\begin{array}
[c]{ll}
\hat{h}_{-} + |\phi_{\rm o}|^{2} & \; \; \; \; \; \; \; \phi_{\rm o}^{2} \\
- \phi_{\rm o}^{\ast 2} & - \left[ \hat{h}_{-} + |\phi_{\rm o}|^{2} \right]
\end{array} 
\right) .
\end{align}
The equations that diagonalize the 
$\hat{\cal L}$ operator,
\begin{equation}
\hat{\cal L} 
\left(
\begin{array}
[c]{l}
u_{j}({\bf x}) \\
v_{j}({\bf x})
\end{array} 
\right) = \omega_{j}
\left(
\begin{array}
[c]{l}
u_{j}({\bf x}) \\
v_{j}({\bf x})
\end{array} 
\right) \; ,
\label{leig}
\end{equation}
are the Bogoliubov-de Gennes (BdG) equations \cite{castinnote}.  

If $\omega_{j}$ is an eigenvalue of $\hat{\cal L}$ then $\omega_{j}^{\ast}$ is also
an eigenvalue.  If $\phi_{\rm o}$ is real-valued, this statement can be verified
by taking the complex conjugate of the BdG equations, but this property is 
true for complex-valued $\phi_{\rm o}$ wavefunctions as well.  Note that if
$\left( \begin{array} [c]{l} u_{j}({\bf x}) \\ v_{j}({\bf x}) \end{array} \right)$ is a
right-eigenvector of $\hat{\cal L}$, as shown in Eq.~(\ref{leig}), then
$\left( u^{\ast}_{j}({\bf x}), - v^{\ast}_{j}({\bf x}) \right)$ is a left-eigenvector of
eigenvalue $\omega^{\ast}_{j}$.  By integrating
\begin{eqnarray}
\int d^{3} x  & \left( u^{\ast}_{j'}({\bf x}), - v^{\ast}_{j'}({\bf x}) \right)
\; \hat{\cal L} 
\left(
\begin{array}
[c]{l}
u_{j}({\bf x}) \\
v_{j}({\bf x})
\end{array} 
\right) 
\nonumber \\
&=
\omega^{\ast}_{j'} \left( \langle u_{j'}| u_{j} \rangle - \langle v_{j'} | v_{j} \rangle \right)
\nonumber \\
&= \omega_{j} \left( \langle u_{j'}| u_{j} \rangle - \langle v_{j'} | v_{j} \rangle \right) \; ,
\end{eqnarray}
where $\langle u_{j'}|u_{j} \rangle = \int d^{3} x\, u^{\ast}_{j'}({\bf x}) u^{\ast}_{j}({\bf x})$,  we
find that if $\omega^{\ast}_{j'} \neq \omega_{j}$, $\langle u_{j'}| 
u_{j} \rangle - \langle v_{j'} | v_{j} \rangle  = 0$.

The above orthogonalization relation suggests the ($\langle u | u \rangle -
\langle v | v \rangle$) form as a scalar product and we could try 
to normalize the $\left( u_{j}, v_{j} \right)$ solutions by requiring $\langle u_{j} |
u_{j} \rangle - \langle v_{j} | v_{j} \rangle = 1$.  This, however is not possible:
if $\left(\begin{array} [c]{l} u_{j,+}({\bf x}) \\ v_{j,+}({\bf x}) \end{array} \right)$ is
a right eigenvector with eigenvalue $\omega_{j}$, then direct substitution shows that
$\left(\begin{array} [c]{l} u_{j,-}({\bf x}) \\ v_{j,-}({\bf x}) \end{array} \right) =
\left(\begin{array} [c]{l} v^{\ast}_{j,+}({\bf x}) \\ u^{\ast}_{j,+}({\bf x}) \end{array} \right)$
is a right eigenvector of eigenvalue $- \omega^{\ast}_{j}$.  If the $_{j,+}$ eigenvector
is normalized by requiring $\langle u_{j,+}| u_{j,+} \rangle - \langle v_{j,+} | v_{j,+} \rangle
= 1$ then $\langle u_{j,-}| u_{j,-} \rangle - \langle v_{j,-} | v_{j,-} \rangle
= - 1$.  It is then convenient to distinguish between a ``$+$'' family, 
$\langle u_{j',+}| u_{j,+} \rangle - \langle v_{j',+} | v_{j,+} \rangle
=\delta_{j',j}$ and a ``$-$'' family with $\langle u_{j',-}| u_{j,-} \rangle - \langle v_{j',-} | v_{j,-} \rangle
= - \delta_{j',j}$.  

We expand the positive and negative frequency components of the wavefunction fluctuations
as
\begin{align}
\left(
\begin{array}
[c]{l}
\delta \phi_{p}({\bf x},t) \\
\delta \phi_{n}({\bf x},t)
\end{array} 
\right) 
= \sum_{j,s} c_{j,s}(t)
\left(
\begin{array}
[c]{l}
u_{j,s}({\bf x}) \\
v_{j,s}({\bf x})
\end{array} 
\right) ,
\label{exp}
\end{align}
where the $s$ subscript indicates the family, $s=+1,-1$, and the
$\left( u_{j,s}, v_{j,s} \right)$ denotes the right eigenvector of $\hat{\cal L}$
with eigenvalue $\omega_{j,s}$.  Substitution of
Eq.~(\ref{exp}) into the linearized Gross-Pitaevskii equation, Eq.~(\ref{lgp2}), yields
\begin{align}
\sum_{k,s'}  i \frac{\partial c_{k,s'}(t) }{\partial t} &
\left(
\begin{array}
[c]{l}
u_{k,s'}({\bf x}) \\
v_{k,s'}({\bf x})
\end{array} 
\right) 
=
\nonumber \\
&  
\sum_{k,s'} \omega_{k,s'} c_{k,s'}(t)
\left(
\begin{array}
[c]{l}
u_{k,s'}({\bf x}) \\
v_{k,s'}({\bf x})
\end{array} 
\right)
\nonumber \\
&+\left(
\begin{array}
[c]{l}
\phi_{\rm o}({\bf x}) \; \delta U_{p}({\bf x},t) \\
-\phi^{\ast}_{\rm o}({\bf x}) \; \delta U_{n}({\bf x},t)
\end{array} 
\right).
\label{lgp3}
\end{align}
Multiplying from the left by $\left( u^{\ast}_{j,s}, -v^{\ast}_{j,s} \right)$ and integrating
over the position coordinate ${\bf x}$ we obtain
\begin{eqnarray}
i \frac{\partial c_{j,s}}{\partial t}
= &&\omega_{j,s} c_{j,s} 
\nonumber \\
&&+s\left[ \int d^{3} x \;
u^{\ast}_{j,s}({\bf x}) \phi_{\rm o} ({\bf x}) \delta U_{p}({\bf x},t) \right.
\nonumber\\
&&+ \left. \int d^{3} x \;
v^{\ast}_{j,s}({\bf x}) \phi^{\ast}_{\rm o} ({\bf x}) \delta U_{n}({\bf x},t) \right].
\end{eqnarray}
Assuming that the perturbation was turned on at a time $t_{0}$, the
solution that satisfies the corresponding boundary condition, 
$c_{j,s}(t') = 0$ when $t'<t_{0}$, is
\begin{eqnarray}
c_{j,s}(t) = && s \int_{t_{0}}^{t} d \tau \, \exp(-i\omega_{j,s}[t-\tau])
\nonumber \\
&&\times \int d^{3} x ' \, 
\left[ u_{j,s}({\bf x}') \phi_{\rm o}({\bf x}') \delta
U_{p}({\bf x}',\tau) \right.
\nonumber \\
&& \left.
+v_{j,s}({\bf x}') \phi^{\ast}_{\rm o}({\bf x}') \delta 
U_{n}({\bf x}',\tau) \right] .
\end{eqnarray}
If $\omega_{j,s}$ has a finite imaginary part, one of the
$c$ amplitudes grows exponentially.  Even if the imaginary
part of the ($j,s$) frequency corresponds to a damped mode, $\omega_{j,s} = \omega - i \Gamma$,
another mode ($n,s$) exists for which $\omega_{n,s}=\omega_{j,s}^{\ast} = \omega + i \Gamma$,
so that $c_{n,s}(t)$ grows exponentially.  In that case, even the smallest of perturbations 
gives rise to an exponentially growing response.  In physical and simulated BECs, the effects of
perturbations are always present, caused either by thermal or quantum fluctuations
in physical systems or by round-off errors in numerical simulations.  From
these considerations the criterion for dynamical stability in linear stability analysis follows:
{\it all} eigenvalues of the BdG equations need to be real-valued for the system
to be dynamically stable.

For studying the stability of the standing wave Gross-Pitaveskii solutions of Section II
(in which case the $\phi_{\rm o}$ wavefunction, and hence $\hat{\cal L}$, is real-valued,
and $\phi_{\rm o}^{2} = |\phi_{\rm o}|^{2}$),  we find it convenient to calculate the sum 
and difference vectors instead of the $u$ and $v$ functions,
\begin{equation}
\label{fpfm}
f^\pm_j=u_j \pm v_j {\rm ,}     
\end{equation}
where, from now on,  the subscripts $j$ fully characterize the mode (including the ``family'').
By adding and subtracting Eq.~(\ref{leig})
the BdG equations take the form of a diagonal set of coupled 
eigenvalue-type equations, 
\begin{eqnarray}
\nonumber
\hat{h}^- f^-_{j} = \omega_j f^+_j  {\rm ;}   \\
\label{bdg2}
\hat{h}^+ f^+_j= \omega_j f^-_j  {\rm ;}
\label{BdGpm}
\end{eqnarray}
where the $\hat{h}^\pm$ operators,
\begin{equation}
\nonumber
\hat{h}^{\pm}= -\frac{\nabla^{2}}{2} + U_{\rm o} + \left[ 2 \phi_{\rm o}^{2} - 1 \right] 
\pm \phi_{\rm o}^{2} \; ,
\end{equation}
are single-particle-Hamiltonian-like operators. 
For a cylindrically symmetric potential and standing wave pattern, the modes will be eigenvectors
of the angular momentum operator.  We perform the separation of variables into radial and angular 
functions,
\begin{equation}
\label{fmodes}
f^\pm_j({\bf x}) = f^\pm_{m,\nu}(\rho)e^{im\theta} {\rm ,}
\end{equation}
where $\rho$ is the dimensionless radial coordinate 
measured in units of the healing length $\xi$,
$\theta$ is the azimuthal angle, and $|m|=0$, $1$, $2$, $\ldots$, 
the azimuthal quantum number.
Henceforth we will take the radial quantum number $\nu$ as 
implicit.  Using Eqs.~(\ref{fpfm}) and (\ref{fmodes}), 
the BdG equations (\ref{BdGpm}) take on the form
\begin{eqnarray}
\nonumber
h^-_m f^-_m = \omega_m f^+_m  {\rm ;}   \\
\label{bdg2}
h^+_m f^+_m = \omega_m f^-_m  {\rm ;}
\end{eqnarray}
where
\begin{eqnarray}
\nonumber
h^+_m = -\frac{1}{2 \rho}\frac{\partial}{\partial \rho} \rho \frac{\partial }{\partial \rho} 
           + \frac{m^2}{2 \rho^2} + U_o(\rho) + 3 \phi_{\rm o}^2 -1  {\rm ;}    \\
\label{hphm}
h^-_m = -\frac{1}{2 \rho}\frac{\partial}{\partial \rho} \rho \frac{\partial }{\partial \rho} 
           + \frac{m^2}{2 \rho^2} + U_o(\rho) + \phi_{\rm o}^2 -1  {\rm ;}   
\end{eqnarray}
denote the radial $\hat{h}^{\pm}$ operators for fixed azimuthal quantum number.
  
What is the physical meaning of the $f^+_m$ and $f^-_m$-fluctuation functions?
At finite oscillation frequency, $f^+_m$ is proportional to the density fluctuation and $f^-_m$ to the phase
fluctuation.  This can be seen by writing the wavefunction as $\phi=\sqrt{n}e^{i\alpha}$,
defining the equilibrium density and phase values as $n_{\rm o}$, $\alpha_{\rm o}$, and introducing
their fluctuations $n=n_{\rm o} + \delta n$, $\alpha = \alpha_{\rm o}+ \delta \alpha$, so that
\begin{equation}
\label{dphi2}
\delta \phi = \frac{\delta n}{2\sqrt{n_{\rm o}}} + i\sqrt{n_o}\delta\alpha  {\rm .}
\end{equation}
We see that we can isolate the density and phase fluctuations by taking 
\begin{eqnarray}
\nonumber
\delta\phi +\delta\phi^*  &=& \frac{\delta n}{\sqrt{n_{\rm o}}}   {\rm ,}   \\
\label{dphi3}
\delta\phi -\delta\phi^*  &=& 2i\sqrt{n_{\rm o}} \delta\alpha  {\rm .}
\end{eqnarray}
But from Eq.~(\ref{phipn}), $\delta \phi = \delta \phi_{p} + \delta \phi^{\ast}_{n}$,
and putting $\delta \phi_{p} = u_{j}$ and $\delta \phi_{n}=v_{j}$ if only the
$j$-mode is excited by the perturbation; we find, with Eqs.~(\ref{fpfm}) and (\ref{fmodes}), 
that 
\begin{eqnarray}
\nonumber
\delta\phi +\delta\phi^*  &=& f^+_m(\rho)e^{im\theta}e^{-i\omega_m t} + {\rm c.c.}   {\rm ,}   \\
\label{dphi4}
\delta\phi -\delta\phi^*  &=& f^-_m(\rho)e^{im\theta}e^{-i\omega_m t} - {\rm c.c.}  {\rm .}
\end{eqnarray}
Comparing Eqs.~(\ref{dphi3}) and (\ref{dphi4}), we obtain
\begin{eqnarray}
\nonumber
\delta n  &=& 2\sqrt{n_{\rm o}} {\rm Re}\Big{[} f^+_m(\rho)e^{im\theta}e^{-i\omega_m t} \Big{]}   {\rm ;}   \\
\label{fluct}
\delta\alpha &=& \frac{1}{\sqrt{n_{\rm o}}} {\rm Im} \Big{[} f^-_m(\rho)e^{im\theta}e^{-i\omega_m t} \Big{]}  {\rm ;}
\end{eqnarray}
so that $f^+_m$ and $f^-_m$ are indeed proportional to the density and phase fluctuations,
respectively. 

The coupled BdG equations, Eqs.~(\ref{bdg2}), can be cast in an alternate form by noting that
the repeated application of $\omega_m$ yields decoupled eigenvalue equations,
\begin{eqnarray}
\nonumber
\omega_m^2 f^+_m = h^-_m h^+_m f^+_m  {\rm ;}   \\
\label{bdg4th}
\omega_m^2 f^-_m = h^+_m h^-_m f^-_m   {\rm ;}
\end{eqnarray}
so that $f^+_m$ is the eigenvector of $h^-_m h^+_m$, whereas $f^-_m$ is the eigenvector of $h^+_m h^-_m$.
While this eigenvalue problem is distinct from the Hamiltonian eigenvalue problem 
($h^\pm_mh^\mp_m$ has a fourth derivative in space, for instance), there is one important similarity:
since the product operators $h^+_mh^-_m$ and $h^-_mh^+_m$ are both self-adjoint, their eigenvalues
$\omega_m^2$ are real-valued.  As a consequence, $\omega_m$ is either entirely real (if $\omega_m^2>0$)
or entirely imaginary (if $\omega_m^2<0$).  The former case, real $\omega_m$, corresponds
to oscillations about a possibly stable equilibrium solution $\phi_{\rm o}$ for mode $m$; 
whereas an imaginary $\omega_m$ indicates that
the mode $m$ drives the system away from an unstable equilibrium solution $\phi_{\rm o}$.

\section{Linear Stability Analysis: 2D BEC Standing Wave Patterns}

We subject the standing-wave $\phi_{\rm o}$ wavefunctions of Section II
to the stability analysis of Section III.  We use the $h_{m}^{\pm}$ form of the 
BdG equations, Eq.~(\ref{hphm}), and  
solve Eqs.~(\ref{bdg4th}) for $\omega_m^2$ and $f^\pm_m$. 
In the following two subsections we describe typical 
examples to uncover the general stability picture.
The first case deals with a shallow well that supports nodal 
BECs with a single node and the second case deals with deeper wells and
nodal BECs with many nodes.  We highlight the following key results:
(i) the nonlinear nodal BECs are always unstable, and (ii) the atomic nodal
BECs can transition from stable to unstable and from unstable
to stable as the well depth increases.

\subsubsection{Single node BEC states in a well of coherence length radius}

We consider the stability of a BEC with one node induced by
a potential well of radius $b=1.0\xi$, as pictured in Fig.~\ref{figwf}.
While the azimuthal quantum number $m$ has to be an integer to 
ensure the single-valuedness of $f_{m}$ in Eq.~(\ref{fmodes}), we can 
formally solve Eqs.~(\ref{bdg2}) for any value of $m$ .  Varying $m$ 
continuously, we plot  ${\rm Im}[\omega_m]$ for the nonlinear (a) and 
atomic (b) nodal BEC solutions with one node for various well depths.

\begin{figure}[htb]
\center{\includegraphics[width=3.3in]{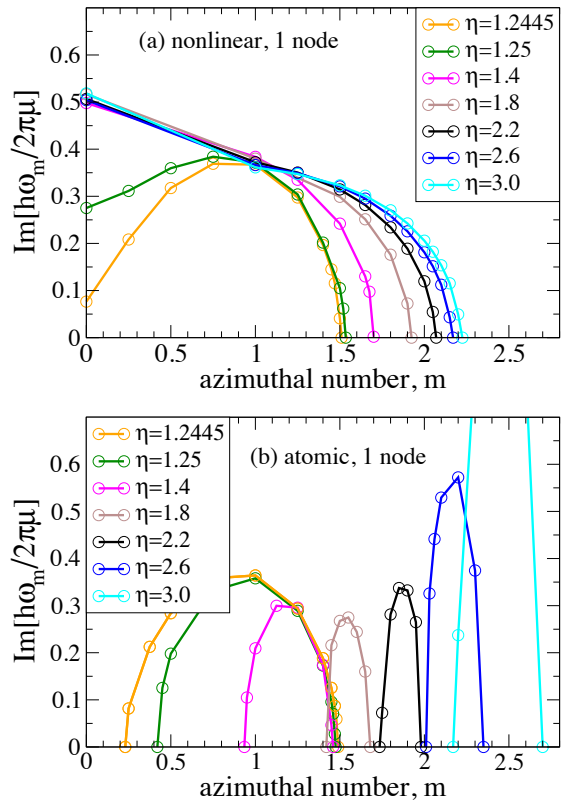}}   
\caption{ \label{figimw} The linear stability spectrum ${\rm Im}[\omega_m]$ is shown for several well depths $\eta$
for (a) the nonlinear nodal BEC and (b) the atomic nodal BEC, both with one node for well radius $b=1.0\,\xi$.  
The noninteger value $m$'s and the connecting lines have been included to guide the eye, although
{\it only integer-valued azimuthal quantum number} m{\it 's are physical}.  
The nonlinear nodal BEC (a) always has at least one
integer $m$ with an imaginary $\omega_m$ and thus is always unstable.  The atomic nodal BEC (b)
has no imaginary-valued $\omega_m$ for integer $m$ when $\eta=1.8$,$\,2.2$,$\,2.6$, and $3.0$,
and is thus stable for these cases.}
\end{figure} 

When the atomic and nonlinear nodal solutions first appear with increasing $\eta$, 
they are identical.  For a well-depth of $\eta=1.25$, we show the two wavefunctions with
one node in Fig.~\ref{figwf}(a).  The blue line shows the atomic nodal BEC wavefunction 
and the red dashed line represents the nonlinear nodal BEC wavefunction.  Note that in spite of their
similarity, the  ${\rm Im}[\omega_m]$ spectra differ markedly, as seen in Fig.~\ref{figimw}.  
The nonlinear nodal BEC has a finite value of ${\rm Im}[\omega_{m=0}]$, while the atomic 
nodal BEC has ${\rm Im}[\omega_{m=0}]=0$, a general features that distinguishes
atomic from nonlinear solutions.   In Fig.~\ref{figimw}, we see that a small variation
of $\eta$ from $\eta=1.2445$ (within $0.0001$ of the critical well depth for nodal solutions)
to $\eta=1.25$, drives the left-edge of the ${\rm Im}[\omega_{m}]$ lobe in opposite 
directions: ${\rm Im}[\omega_{m=0}]=0$ for the critical well depth where the nodal solutions
first exist; and $\omega_{m=0}$ becomes entirely imaginary for the nonlinear nodal 
BEC solutions and entirely real for the atomic nodal BEC solutions.  
The nonlinear nodal BECs have at least one imaginary-valued frequency mode, 
the radially symmetric $m=0$ mode, so that the nonlinear nodal BECs are always unstable.  The
instability of the nonlinear nodal BECs has features in common with the snake instability 
of a kinkwise nodal line or plane in an otherwise homogeneous 2D or 3D condensate 
\cite{muryshev}, but an in-depth discussion would take us further afield.

We now discuss the atomic nodal solutions.  Fig.~\ref{figimw}(b) shows ${\rm Im}[\omega_m]$ 
for the one-node atomic nodal BEC state.  As mentioned above, ${\rm Im}[\omega_{m=0}]=0$
and the lowest possible angular momentum of an exponentially growing mode is
$m\ge1$.  For the lowest three well depth values shown in Fig.~\ref{figimw}(b), the $m=1$ mode
has an imaginary-valued frequency and the atomic nodal BEC is unstable.  As the well 
increases, from  $\eta=1.8$ to $\eta=2.2$, for instance, we don't find any imaginary $\omega_m$ 
values for integer $m$ and the atomic nodal BEC is stable.  Between $\eta=2.2$ and $\eta=2.6$, 
the ${\rm Im}[\omega_m]$ lobe shifts to larger $m$ values and passes through the $m=2$ line, indicating
the instability of 
the mode with azimuthal quantum number 2.  In the $\eta=2.6$ to $\eta=3.0$ interval, 
the ${\rm Im}[\omega_m]$ lobe has shifted to values between $m=2$ and $m=3$, 
so that this regime of well-depths is stable again.  
Therefore, as the potential deepens from $\eta=1.25$ to $3.0$, the single node atomic
nodal BEC transitions from instability with exponentially growing $m=1$ mode to stability, then
returns to an unstable status, this time caused by the exponential growth of the $m=2$ mode, 
then enters another island of stability.  We refer to this trend as {\it reentrant stability} to
describe the entering and exiting into and out of islands of stability.

\subsubsection{Multi-node BEC states in broad, deep wells}

\begin{figure}[htb]
\center{\includegraphics[width=3.0in]{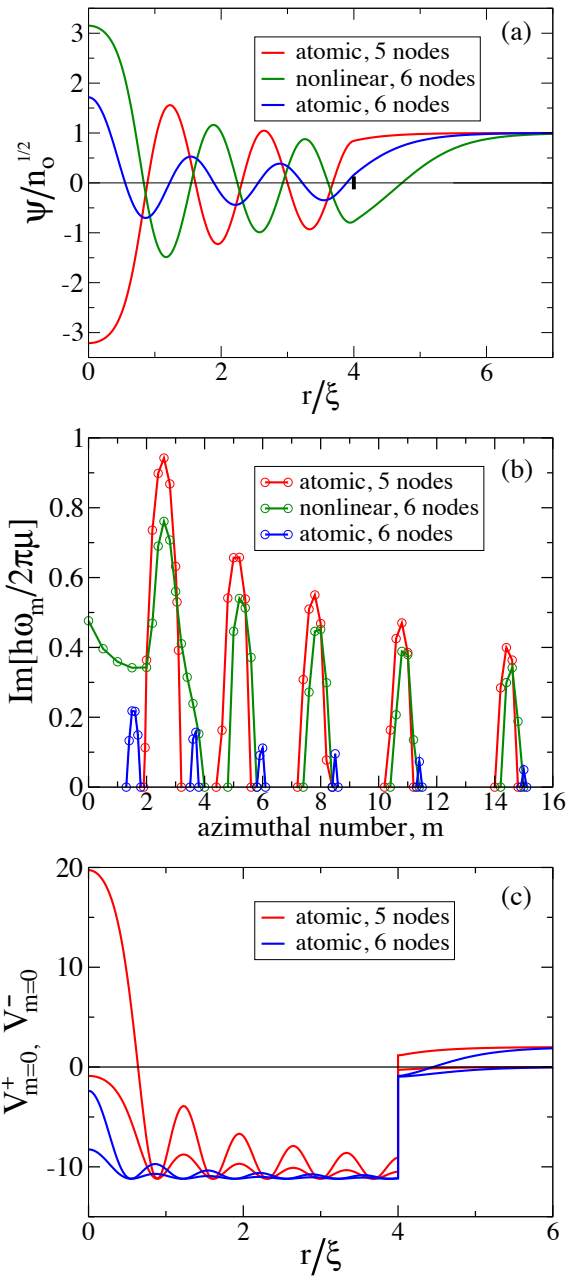}} 
\caption{ \label{figimw2}  (a) Nodal BEC wavefunctions five and six
nodes are shown for a well radius $b=4.0\,\xi$ and well depth
$\eta=11.5$. (b) The linear stability spectrum ${\rm Im}[\omega_m]$ for the three solutions
of (a); as in Fig.~\ref{figimw}, only integer $m$'s are physical and the noninteger $m$'s and lines
are shown to clarify the trends. 
(c) The potentials $V^\pm_{m=0}$ of Eq.~(\ref{vpvm})   
are useful for interpreting the linear stability spectra, as discussed in the text.  }
\end{figure}

We now discuss nodal BECs in a wider, deeper well with radius $b=4.0\,\xi$ and depth 
$\eta=11.5$.  This well has nodal BEC solutions with up to six nodes; Fig.~\ref{figimw2}(a) 
shows the five node
atomic and six node atomic and nonlinear nodal BECs.  Fig.~\ref{figimw2}(b) shows the 
corresponding ${\rm Im}[\omega_m]$ spectra. As in the previous subsection, we find 
that the nonlinear solution has an imaginary-valued $m=0$ mode, while the two atomic 
solutions do not, following the general $m=0$ behavior mentioned above.
More striking, however, are the spectra's multi-peaked structures which we now explain.

To understand the ${\rm Im}[\omega_m]$ structure we revisit the BdG equations 
(\ref{bdg2}), $h^\pm_m f^\pm_m = \omega_m f^\mp_m$.  The frequency $\omega_m$ 
is either entirely real or entirely imaginary.  Therefore, if we treat $m$ as a continuous variable
and we assume $\omega_m$ to be a continuous function of $m$ then
$\omega_m$ has to pass through zero to switch from real to imaginary values.
When $\omega_m$ vanishes, either $h^+_mf^+_m=0$ or $h^-_mf^-_m=0$,
corresponding to a $f^+_m$ or $f^-_m$ eigenvector of the $h^{\pm}_{m}$ operator with
zero eigenvalue.  

Consider the $h^\pm_m$ operators of (\ref{hphm}) which we write as
$h^\pm_m=-\frac{1}{2\rho}\frac{\partial}{\partial\rho}\rho\frac{\partial}{\partial\rho} + V^\pm_m$
introducing the $V^\pm$ potentials,
\begin{equation}
\label{vpvm}
V^\pm_m=\frac{m^2}{2\rho^2}+U_o(\rho) +\phi_o^2(2\pm 1)-1\; ,
\end{equation}
that are effective potentials of the single-particle-Hamiltonian-like operators $h^{\pm}_{m}$.
We show examples of $V^\pm_m$ for $m=0$ in Fig.~\ref{figimw2}(c) 
for the two atomic nodal BECs of Fig.~\ref{figimw2}(a).  The short-scale,
oscillatory structure of the $V^{\pm}$ potentials stems from the $\phi_{\rm{o}}$ contributions,
but for high-node BEC states in sufficiently deep wells (for which $\phi^{2}_{\rm o}\ll U_{\rm o}$)
the overall $V^{\pm}$ shape is dominated by the external $U_{\rm o}$ potential. 
For example, Fig.~\ref{figimw2}(c) shows $V^{\pm}$ for
the state with five (six) nodes and the $V^{\pm}$ potentials themselves support five (six)
negative energy levels.  As $m$ increases, the angular momentum barrier, $m^{2}/2\rho^{2}$,
grows and raises the energy-levels to positive values \cite{notevpvm}.  At a value $m_1$, 
a $V^+_{m_1}$ level crosses the zero energy dividing line and $h^+_{m_1}f^+_{m_1}=0$ 
so that the $h^+_m$ BdG equation becomes an eigenvalue equation with zero eigenvalue.  
This $m_1$ value denotes the left-hand side of the first peak or lobe in the ${\rm Im}[\omega_m]$
spectrum.  As $m$ increases further to $m_2$ at which the corresponding $V^-_{m_2}$ level
crosses over, $h^-_{m_2}f^-_{m_2}=0$, the $h^-_m$ BdG equation reduces to another
eigenvalue equation with zero eigenvalue and the $m_{2}$ value denotes the right-hand
edge of the lobe that started at $m_{1}$.  The pattern repeats as many times as there
are levels in the $V^{\pm}_{m=0}$ potentials, generating the other peaks in the 
${\rm Im}[\omega_m]$ spectrum.

The above construction shows that the number of peaks in the ${\rm Im}[\omega_m]$ spectrum  
is equal to the number of negative energy levels of the $V^\pm_{m=0}$ potentials.  Moreover, 
the widths of the ${\rm Im}[\omega_m]$ peaks, $\Delta m$, are related to the
energy splitting of corresponding states in the $V^{+}_{m=0}$ and in the
$V^{-}_{m=0}$ potentials.  We illustrate this point with Fig.~\ref{figimw2}(c): as
the BEC density $\phi_{\rm o}^{2}$ in the potential well is smaller in the six node 
than in the five node BEC, the $V^{+}_{m=0}$ and $V^{-}_{m=0}$ potentials 
resemble each other more in the six node atomic BEC state than in the five node 
atomic BEC state.  As a consequence, the difference in energy between
corresponding pairs of $V^{+}_{m=0}$ and $V^{-}_{m=0}$ eigenstates is smaller 
in the six node than in the five node case.  Therefore, the $m$ value at which 
the $V^{+}$ and the $V^{-}$ state reaches zero energy differ by a smaller
amount $\Delta m$ in the six node than in the five node situation.  Generally,
as the atomic nodal BEC with the maximal number of nodes has the smallest BEC density
$\phi_o^2$, this time-independent BEC solution has the smallest peak widths $\Delta m$.

The above analysis reveals that common methods for bound state physics can
determine relevant instability properties of standing wave BECs.  For instance, we
consider the upper imaginary azimuthal quantum number $m_{c}$ above which
all mode eigenfrequencies $\omega_{m}$ take on purely real values.  We estimate
$m_{c}$ for the atomic nodal BEC state of highest node number contained in a
potential well of sufficient depth to ensure that the BEC density 
satisfies $|\phi_o|^2\ll U_o$ inside the well (for the nodeless case, this is obviously not true).  
As we ramp up the $m$ quantum number, the last bound state of the $V^{-}_{m}$ potential 
becomes unbound at $m=m_{c}$.  The corresponding $V^{-}_{m}$ eigenstate 
(resonant) wavefunction, which we shall refer to as $\chi^{-}_{\rm{o}}$ (and we assume 
$\chi^{-}_{\rm{o}}$ to be normalized) is a nodeless function roughly centered on the
position of the potential minimum of $V^{-}_{m}$, i.e., on $\rho =b$.  At the zero energy
crossing point,
\begin{equation}
\langle \chi^{-}_{\rm{o}} | - \frac{\nabla^{2}}{2} + U_{\rm{o}} + \frac{m_{c}^{2}}{2\rho^{2}}
+\left[ \phi_{\rm{o}}^{2}-1 \right] | \chi^{-}_{\rm{o}} \rangle = 0
\label{cross}
\end{equation}
Neglecting the kinetic energy term and the $\left[ \phi_{\rm{o}}^{2}-1 \right]$ contributions
and estimating the angular momentum barrier contribution by $m_{c}^{2}/2b^{2}$,
we find $m_{c} \approx b\sqrt{2 U_{\rm{o}}}$, or with our notation for the square well potential
\begin{equation}
\label{mc}
m_c \approx \frac{\pi \eta}{2},
\end{equation}
which overestimates $m_c$ because we neglected the kinetic energy and the
$\phi_o^2$ terms.  Nonetheless, we expect this approximation to be reasonable for deep wells.  
For example, in Fig.~\ref{figimw2}, the observed $m_c$ of $\sim 15$
can be compared to the $m_c$ estimate of $18.1$ from Eq.~(\ref{mc}).

\begin{figure}[htb]
\center{\includegraphics[width=3.0in]{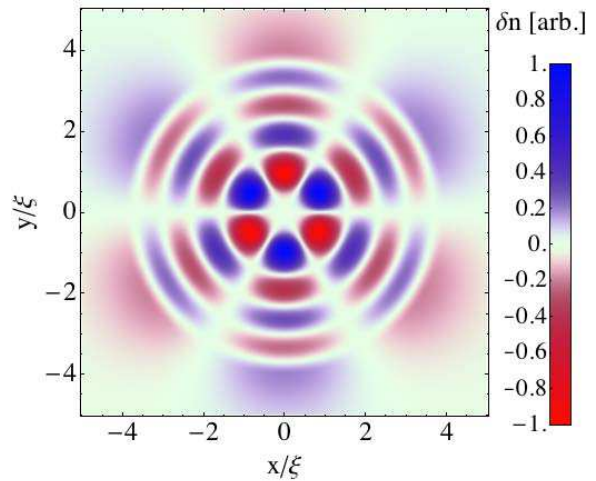}} 
\caption{ \label{figdensityfluc} The density fluctuation for the $m=3$ mode, $\delta n$ from Eq.~(\ref{fluct}),
is shown for the five node atomic nodal BEC state shown in Fig.~\ref{figimw2}(a).  This is an unstable
mode with an imaginary $\omega_m$ and the pattern seen here will contribute to the break-up of this nodal BEC.}
\end{figure}

All three nodal BEC-states shown in Fig.~\ref{figimw2} are unstable.  The five node atomic 
nodal BEC has imaginary modes for $m=2$,$\,3$,$\,5$,$\,8$,$\,11$, and $14$.  In 
Fig.~\ref{figdensityfluc} we show the density fluctuation pattern of the $m=3$ mode, proportional
to $f^+_m(\rho)\cos(m\theta)$ in accordance with Eq.~(\ref{fluct}).  If the $m=3$ mode were
the only unstable mode or if it were the dominant unstable mode (growing at the largest rate 
$\tau_{m}^{-1} = \rm{Im}(\omega_{m})$), the density variation would take on the
form pictured in Fig.~\ref{figimw2} in the initial stage ($t<\tau_{m}$) of the break-up
of the nodal BEC state.  The intricate oscillatory pattern of Fig.~\ref{figdensityfluc} highlights the
role of interference in the wave dynamics of the instability.  Only when
the interference fringes ``match up'' properly to initiate the break-up does the instability
set in.  In addition to the restriction of $m$ values to integers, which is an expression of the
angular boundary condition $\delta \phi(\rho,\theta)=\delta \phi(\rho,\theta+2\pi)$,
the radial oscillations have to line up to create the conditions at which the instability
can set in.  We interpret the finite $\Delta m$ width of the ${\rm Im}[\omega_m]$ lobes
as the range of $m$ values for which the radial interference fringes 
match up sufficiently well to trigger the break-up of the nodal BEC structure. 

The six node atomic nodal BEC whose $\rm{Im}[\omega_{m}]$ spectrum is shown 
in Fig.~\ref{figimw2} has unstable modes for $m=6$ and $m=15$. However, the 
widths of its $\rm{Im}[\omega_{m}]$ lobes satisfy $\Delta m <1$.  This means that as the depth 
of the potential well is varied, the imaginary-valued $\omega_m$ intervals can shift off 
of all integer $m$'s leading to a reentrant stability behavior similar to that discussed 
in the previous subsection.

Can we make a general statement about the stability of stationary multi-node BEC wavefunctions?
The multi-lobed structure of the $\rm{Im}[\omega_{m}]$ spectrum can actually promote
stability: if the unstable $m$ intervals, $\Delta m$, are sufficiently narrow, 
$\Delta m \ll 1$, the $\rm{Im}[\omega_{m}]$ lobes may not overlap any integer
$m$ values.  The progression from the five node to the six node $\rm{Im}[\omega_{m}]$ spectrum
shown in Fig.~\ref{figimw2}(b) indicates an interesting trend: while the increase in nodes
increases the number of lobes, their $\Delta m$ widths decrease markedly.
Furthermore, the $\Delta m$ widths decrease for wells of smaller radius, as we now show.

We can derive a rough estimate for the peak widths $\Delta m$ along the lines of the above
argument that associates the vanishing of $\omega_m$ with the zero-energy crossing
of bound state levels of $V^\pm_m$.  We assume that the BEC density inside the
well remains sufficiently small to determine the splitting between the $V^{+}_{m}$ and 
$V^{-}_{m}$ eigen energies by perturbation theory.  The necessary condition reduces
to $\phi^{2}_{\rm{o}}\ll |U_{\rm{o}}-1|$ in the relevant well region---the region in which the
the $V^{\pm}_{m}$ eigenstates have significant density.
We determine the $V^{+}_{m}$ and $V^{-}_{m}$ eigenstate energies as perturbations
on the eigenstate of the average potential $\overline{V}_{m}=\left[ V^{+}_{m}
+ V^{-}_{m} \right]/2$,
\begin{equation}
\overline{V}_{m}(\rho) 
= U_{\rm{o}}(\rho) + 2 \phi^{2}_{\rm{o}}(\rho) - 1 + \frac{m^{2}}{2 \rho^{2}} \; .
\label{overlinev}
\end{equation}
We denote the eigenstate of the $\overline{V}_{m}$ potential with $\nu$ nodes
by $\chi_{m,\nu}$ and we denote its energy by $\overline{E}_{\nu}(m)$.  In first order perturbation
theory, the energy of the corresponding eigenstates of the $V^{+}_{m}$ and 
$V^{-}_{m}$ potentials, $E_{\nu}^{+}(m)$ and $E_{\nu}^{-}(m)$, are then equal to
\begin{equation}
E_{\nu}^{\pm}(m) = \overline{E}_{\nu}(m) \pm \Delta E_{\nu}(m) \; ,
\label{pert}
\end{equation}
where $\Delta E_{\nu}(m)$ is the first-order $\phi^{2}_{\rm{o}}$ energy splitting,
\begin{equation}
\Delta  E_{\nu}(m) = \langle \chi_{m,\nu}| \phi^{2}_{\rm{o}} | \chi_{m,\nu} \rangle \; .
\label{pert2}
\end{equation}
The slope of the $m$ variation of $\overline{E}_{\nu}(m)$, $\partial \overline{E}_{\nu}/
\partial m$, can be calculated with the Hellmann-Feynman theorem
(the derivative of the hamiltonian expectation value with respect to a
hamiltonian parameter is the expectation value of the derivative of the 
hamiltonian operator):
\begin{eqnarray}
\frac{\partial \overline{E}_{\nu}}{\partial m}
&=& \langle \chi_{m,\nu} | \frac{\partial \overline{V}_{m}}{\partial m} | \chi_{m,\nu}\rangle
\nonumber \\
&=& \langle \chi_{m,\nu} | \frac{m}{\rho^{2}} | \chi_{m,\nu} \rangle \; .
\label{dedm}
\end{eqnarray}
Then, the difference $\Delta m_\nu$ in $m$ values at which the $E_{m}^{+}$ and the $E_{m}^{-}$ values
cross the zero-energy line can be calculated from the matrix elements for
$m=\overline{m}_{\rm{o},\nu}$ at which the $\nu$-level of the average 
$\overline{V}_m$ potential crosses the zero energy line,
\begin{eqnarray}
\Delta m_{\nu} &=&
\frac{ 2 \Delta E_{\nu}(\overline{m}_{\rm{o},\nu}) 
}
{ 
\partial \overline{E}_{\nu}/\partial m  }
\nonumber \\
&=& \frac{ 
\langle \chi_{\overline{m}_{\rm{o},\nu}} | \phi^{2}_{\rm{o}} | 
\chi_{\overline{m}_{\rm{o},\nu}} \rangle
}
{ 
\overline{m}_{\rm{o},\nu} \langle \chi_{\overline{m}_{\rm{o},\nu}} | \rho^{-2} | 
\chi_{\overline{m}_{\rm{o},\nu}} \rangle
} \; ,
\label{dm}
\end{eqnarray}
Assuming that the $\chi_{\overline{m}_{\rm{o},\nu}}$ function is centered on
the potential edge $\rho=b$, we can estimate the denominator as 
$\overline{m}_{\rm{o},\nu} /b^{2}$.  Denoting $\langle \chi_{\overline{m}_{\rm{o},\nu}}
| \phi_{\rm{o}}^{2} | \chi_{\overline{m}_{\rm{o},\nu}} \rangle$ by 
$\left(\overline{\phi_{\rm{o}}^{2}}\right)_{\overline{m}_{\rm{o},\nu}}$, we find
\begin{equation}
\Delta m_{\nu} \approx \frac{2 b^{2}}{\overline{m}_{\rm{o},\nu}} 
\left(\overline{\phi_{\rm{o}}^{2}}\right)_{\overline{m}_{\rm{o},\nu}} \; .
\end{equation}
Since the $\chi_{\overline{m}_{\rm{o},\nu}}$ function is centered on the edge
of the $U_{\rm{o}}$ potential where the $\phi_{\rm{o}}$ function is reaching
towards its asymptotic value, we try an even cruder approximation $\phi_{\rm{o}}^{2} \rightarrow 1$,
resulting in 
\begin{equation}
\Delta m_{\nu} \approx \frac{2 b^{2}}{\overline{m}_{o,\nu}} \; ,
\label{deltam}
\end{equation}
which overestimates the $\Delta m$ widths. Nevertheless, we believe that the general trends 
predicted by Eq.~(\ref{deltam}) are correct.

From Eq.~(\ref{deltam}), we expect that the peak widths $\Delta m$ of the 
${\rm Im}[\omega_m]$ spectrum narrows (i) as the lobe is situated at larger 
$m$ values, and (ii) as the radius of the potential well decreases.
The former effect can be observed in the spectrum of Fig.~\ref{figimw2}(b).  
The latter effect implies that narrow wells with a subcoherence-length radius,
$b\ll\xi$, have peaks of widths $\Delta m\ll1$, implying a decreased likelihood 
of lobe-overlap with an integer $m$ value (hence, increased likelihood of stability).

\section{Dynamical Stability from Single-particle Hamiltonian-like
Eigenspectra: Recipe}

The treatment of Section IV makes explicit use of the
cylindrical symmetry by introducing the device of a continuously
varying azimuthal quantum number $m$.  In this section, we propose
a different, more general procedure to test the dynamical stability
of localized, potential-induced standing wave BEC patterns.
The recipe we describe below is based on the same concept 
of continuously varying frequencies $\omega_{j}$ that
have to vanish as they cross over from real to imaginary
values.  At the crossing points, the BdG
equations reduce to a single-particle-Hamiltonian-like
eigenproblem for zero eigenvalue.  As in Section IV, the general
recipe allows one to infer the stability of a standing wave
pattern from the eigenspectra of single-particle-like
operators alone, circumventing the need of conducting the
full linear stability analysis.  Unlike the technique of Section
IV, the recipe is independent of the symmetry of the potential
and of the standing wave density $|\phi_{\rm o}|^{2}$ and can
be applied in three as well as two dimensions.

\underline{Recipe}: We consider the eigenspectrum of the 
single-particle-Hamiltonian-like operators
\begin{equation}
\label{he}
\hat{h}^{\pm} =
- \frac{\nabla^{2}}{2} + U_{\rm o}({\bf x}) + 2 \left( \phi_{\rm o}^{2} -1 \right)
\pm \phi_{\rm o}^{2} \; .
\end{equation}
We identify pairs of ``similar'' eigenstates $j$ of the $\hat{h}^{+}$ and
$\hat{h}^{-}$ operators with eigenvalues $E^{+}_{j}$ and $E^{-}_{j}$ 
respectively and with the same quantum number $j$, indicating 
eigenstates that would go over into each other in the limit that 
$\phi_{\rm o} \rightarrow 0$ in Eq.~(\ref{he}). We suggest that the standing 
wave solution $\phi_{\rm o}$ of the time-independent Gross-Pitaevskii 
equation is unstable if there is one or more quantum numbers $j$ for 
which $E^{-}_{j} < 0$ and $E^{+}_{j} > 0$ and that it is stable otherwise.

\underline{Justification}:
The justification of the recipe is based on the structure of the BdG 
equations and the continuity of its eigenfrequencies $\omega_{j}$
with respect to a continuously varying parameter $\epsilon$ that appears 
in the Hamiltonian.  Such parameter could be introduced in different
ways (multiplying the chemical potential or the interaction strength, for
instance).  Here we choose to vary the depth of the external potential
\begin{equation}
U_{\epsilon}({\bf x}) = \epsilon U_{\rm{o}} ({\bf x}),
\end{equation}
and we let $\epsilon$ range from $0$ to $1$, assuming that a standing wave 
solution has appeared by $\epsilon=\epsilon_{c}$,
with $\epsilon_{c}\leq1$.  The standing wave function $\phi_{\rm{o},\epsilon}$
solves the time-independent Gross-Pitaevskii equation
\begin{equation}
\left[ - \frac{\nabla^{2}}{2} + U_{\epsilon}({\bf x}) + 
|\phi_{\rm{o}, \epsilon}({\bf x})|^{2} - 1 \right] \phi_{\rm{o}, \epsilon}({\bf x}) = 0
\end{equation}
with $\epsilon > \epsilon_{c}$.  The BdG
equations for fixed $\epsilon$ and real-valued $\phi_{\rm o}$ function,
\begin{eqnarray}
\hat{h}^{+}_{\epsilon} f^{+}_{j,\epsilon} &=& \omega_{j} (\epsilon) f^{-}_{j,\epsilon} ;
\nonumber \\
\hat{h}^{-}_{\epsilon} f^{-}_{j,\epsilon} &=& \omega_{j} (\epsilon) f^{+}_{j,\epsilon} ;
\label{bdg4}
\end{eqnarray}
with
\begin{equation}
\hat{h}^{\pm}_{\epsilon} =
- \frac{\nabla^{2}}{2} + U_{\epsilon}({\bf x}) + 2 \phi_{\rm{o},\epsilon}({\bf x})^{2} - 1
\pm \phi_{\rm{o},\epsilon}^{2}({\bf x}) \; ,
\end{equation}
has solutions that are pairs of functions $( f^{+}_{j,\epsilon},f^{-}_{j,\epsilon})$.
As an eigenvalue of the Hermitian $\hat{h}^{+}_{\epsilon} \hat{h}^{-}_{\epsilon}$ operator,
$\omega_{j}^{2}(\epsilon)$ is real-valued so that $\omega_{j}(\epsilon)$ is either entirely real or
entirely imaginary.  To cross from real to imaginary values, $\omega_{j}({\epsilon})$
has to pass through zero, at which point the BdG equations
reduce to a Hamiltonian eigenvalue problem of vanishing eigenvalue.
If $\chi_{j}^{-}$ is the zero-eigenvalue, the $j$-type eigenvector of the 
$\hat{h}^{-}_{\epsilon}$ operator for $\epsilon=\epsilon^{-}_{j}$,
\begin{equation}
\hat{h}^{-}_{\epsilon^{-}_{j}} \chi_{j}^{-}=0
\end{equation}
then $(f^{+}_{j}=0,f^{-}_{j}=\chi_{j}^{-})$ solves the BdG
equations for that specific $\epsilon$ value.  Conversely, if $\chi_{j}^{+}$ is 
the zero-eigenvalue, $j$-type eigenvector of  the $\hat{h}^{+}_{\epsilon}$ operator
for $\epsilon=\epsilon^{+}_{j}$,
\begin{equation}
\hat{h}^{+}_{\epsilon^{+}_{j}} \chi_{j}^{+}=0
\end{equation}
then $(f^{+}_{j}=\chi_{j}^{+},f^{-}_{j}=0)$ solves the BdG
equations for $\epsilon=\epsilon^{+}_{j}$. In general, $\hat{h}^{\pm}_{\epsilon} \chi_{j,\epsilon}
= E_{j}^{\pm} (\epsilon) \chi_{j,\epsilon}^{\pm}$ and the zero eigen energy crossings happen
at specific $\epsilon$-values, $\epsilon=\epsilon^{\pm}_{j}$ for which 
$E^{\pm}_{j}(\epsilon^{\pm}_{j})=0$.  
As $(0,\chi^{-}_{j})$ and
$(\chi^{+}_{j},0)$ solve the BdG equations for specific
values of the parameter $\epsilon$, the $f^{\pm}_j$ functions must 
be of the same type (same number of nodes, same quantum numbers) as 
the $\chi^{\pm}_j$ wavefunctions.  As $\omega_{j}(\epsilon)$ passes through
zero at $\epsilon=\epsilon^{\pm}_{j}$, it either changes sign or switches between
real and imaginary values.  
We argue for the latter: as an eigenvalue of the $\hat{h}^{+}_{\epsilon} \hat{h}^{-}_{\epsilon}$
operator, $\omega^2_j(\epsilon)$ and its derivative with respect to $\epsilon$ are continuous.
As a consequence, $\omega^2_j(\epsilon)$ switches sign at $\epsilon = \epsilon^{\pm}_j$,
corresponding to $\omega_j(\epsilon)$ switching between real and imaginary values.
If $\epsilon^{-}_{j }< 1$ so that the $E_{j}^{-}(\epsilon)$
eigenvalue has already been lowered through the zero-energy dividing
line by $\epsilon=1$ whereas $\epsilon^{+}_{j} > 1$ so that $E_{j}^{+}>0$
and the $\hat{h}^{+}_j$ eigenvalue has not been brought below
the zero-energy line yet, then $\omega_{j}$ has moved through $0$
only once, switching its value from real to imaginary value.  If both
$E_{j}^{-} < 0$ and $E_{j}^{+} < 0$, then $\omega_{j}(\epsilon)$ moved
twice through $0$ as $\epsilon$ varied from $\epsilon_{c}$ to $1$
switching from real to imaginary and back to real.
The upshot is that if both eigenvalues $E_{j}^{\pm}$ find themselves
on the same side of zero, then $\omega_{j}$ is real, if they are on
opposite sides of the zero energy dividing line, $\omega_{j}$ is
imaginary.

The application of the above method to the 2D-cylindrically symmetric
situation described in Sections II and IV leads to the investigation of
$E_{m}^{\pm}$ for integer, fixed values of $m=0,\pm1,\pm2,...$, from
which stability or instability can be determined.  In contrast, the treatment
of Section IV used the quantum number $m$ itself as the continuously
varying parameter and calculated the rate of the instability ${\rm Im}[\omega_{m}]^{-1}$
as a function of $m$.  A single ${\rm Im}[\omega_{m}]$ curve then reveals 
excitations of different $j$-types: the different ${\rm Im}[\omega_{m}]$-lobes shown
in Fig.~(\ref{figimw2}) correspond to $f$-functions of different $j$-types, here modes
of different vibrational quantum numbers $\nu$ (with different numbers of nodes).
While convenient for studying the 2D cylindrically symmetric standing
wave pattern, the method of a continuously varying quantum number 
cannot readily be generalized to describe non-symmetric $\phi_{\rm{o}}$-patterns
or three-dimensional structures; but the concept outlined here of varying a
parameter in the effective Hamiltonian can be.

\section{Conclusions}

We have studied the existence and dynamical stability of stationary BEC states with nodes 
in the BEC wavefunction.  In particular, we investigated large 2D BEC systems in the presence of an
attractive potential well with a  coherence length-sized radius.  The potential is cylindrically
symmetric and of the square well type.  From the center of the potential well, the BEC
tends to a uniform density.  As the depth of the attractive well increases, BEC solutions 
with radial nodes in the wavefunction appear in pairs, each pair exhibiting one more node 
than the last pair. While the calculations are carried out for this very specific case, we
discuss which trends are independent of the symmetry and can be applied to three-dimensional
standing wave BEC patterns.

We have classified the nodal BEC solutions to the time-independent Gross-Pitaevskii-equation
to be of two types.  One type we refer to as the atomic nodal BEC state since the wavefunction
resembles that of single-particle-like bound states in the attractive potential well. The other type
we refer to as the nonlinear BEC state since the $n$ node BEC wavefunction resembles the
$n-1$ node atomic BEC wavefunction with a soliton attached outside the potential well.

Our stability analysis shows that the nonlinear nodal BEC state is always unstable
and that atomic nodal BEC states can transition from stable to unstable to stable
status as the attractive well depth is adiabatically increased---a behavior which 
we refer to as reentrant stability.  In general, the stable regimes grow
as the radius of the attractive potential is decreased below the BEC healing length scale.

We describe a more general analysis of the dynamical instability that is based
on the eigenspectra of two effective single-particle-Hamiltonian-like operators, which
gives new insights, can be applied to low symmetry standing wave patterns 
patterns, and is valid in three dimensions.

Given the dynamical stability in specific parameter regimes, atomic nodal BEC states
could find applications, for instance, in a BEC ring as a rotation sensor.  Nearby instabilities
may be exploited as a threshold measurement tool.

This work serves as an introduction to the existence and dynamical stability of nodal BEC states.
Many challenges, on the theory and on the experimental front, remain.
We have shown that nodal BEC states can be dynamically stable, but the
manufacturing of these states may require creativity.  Perhaps the phase imprinting
techniques that have generated vortices and dark solitons \cite{phase1,phase2} can
create the required initial conditions.  
Conversely, the non-adiabatic dynamics of BECs in steep localized potential wells may
exhibit signatures of the instabilities described above: if the BEC dynamics brings
the system close to a nodal wavefunction, it can break up if the potential depth is such
that the nodal state is unstable.

We believe that standing wave BECs with nodal surfaces separated by macroscopic 
scale distances suggest new prospects and we hope that this work will stimulate 
interest in standing wave BECs.

\section{Acknowledgments}

This work was funded by the Los Alamos Laboratory Directed Research and Development
(LDRD) program.  All authors thank Malcolm Boshier for interesting conversations and 
the description of dynamical BEC experiments that directly motivated this work.  
R.M.K. and E.T. thank Fernando Cucchietti for useful discussions on 
the integration of the time-independent Gross Pitaevskii equation for steep 3D potentials.


\begin{thebibliography}{10}
\bibitem{remark1} Quantum fluctuations will give a finite density at the nodal
surfaces, so that a more correct statement would refer to a ``nearly vanishing
density''.
\bibitem{notenodej} The existence of nodal BEC solutions with $j$ nodes at well depths shallower
than that necessary for the existence of the noninteracting single particle bound state with $j$ nodes
is due to the presence of the BEC chemical potential creating an effectively deeper well.
\bibitem{notedfpos} The reason that $\Delta F>0$ for well depths near those where the nodal solutions
first exist is related to the appearance of the nodal BEC solutions at well depths shallower
than that for the corresponding non-interacting bound states, cf.~\cite{notenodej}.
\bibitem{castinnote} The $\hat{\cal L}$ operator is actually not quite diagonalizable, as discussed in \cite{castin2}.
This problem can be remedied by introducing a projection operator that projects out the overlap of $\delta\phi$
with $\phi_o$.  However, the modified $\hat{\cal L}$ operator has the same spectrum as the $\hat{\cal L}$ 
operator, so the conclusions remain unaltered.
\bibitem{notevpvm} $V^-_m$ goes to $0$ for large distances and $V^+_m$ goes to $2$ since $\phi_o^2\rightarrow 1$;
thus the $V^-_m$ energy level passing through $0$ corresponds to an unbinding transition while for
$V^+_m$ it does not.
\end{thebibliography}
\end{document}